%% file: ambient.tex
\documentclass{IEEEcsmag}

\usepackage[colorlinks,urlcolor=blue,linkcolor=blue,citecolor=blue]{hyperref}
\expandafter\def\expandafter\UrlBreaks\expandafter{\UrlBreaks\do\/\do\*\do\-\do\~\do\'\do\"\do\-}

\usepackage{mdframed}

\newcommand{\hl}[1]{#1}

\usepackage{multibib}
\newcites{bg}{Further Reading}
\makeatletter
\let\orig@citebg\citebg
\renewcommand{\citebg}[1]{%
  \begingroup
  \def\@cite##1##2{[B##1\if@tempswa , ##2\fi]}%
  \orig@citebg{#1}%
  \endgroup
}
\makeatother
\input{commands}

\newcommand{\papertitle}{Ambient Analytics: Calm Technology for Immersive Visualization and Sensemaking}

\jname{IEEE Computer Graphics \& Applications}
\jtitle{\papertitle}
\pubyear{2026}

\begin{document}

%% ---------------------------------------------------------------------
%% Title, author(s), and affiliations
%% ---------------------------------------------------------------------

%% Magazine department
\sptitle{DEPARTMENT: VISUALIZATION VIEWPOINTS (AUTHOR VERSION)}

%% Paper title
\title{\papertitle}

%% Authors and affiliations
\author{Sebastian Hubenschmid}%
\affil{Aarhus University, Aarhus, Denmark}

\author{Arvind Srinivasan}%
\affil{Aarhus University, Aarhus, Denmark}

\author{Niklas Elmqvist}%
\affil{Aarhus University, Aarhus, Denmark}

\author{Dieter Schmalstieg}%
\affil{University of Stuttgart, Stuttgart, Germany and Graz University of Technology, Graz, Austria}

\author{Michael Sedlmair}%
\affil{University of Stuttgart, Stuttgart, Germany}

%% Running headings
\markboth{SPECIAL ISSUE}{SPECIAL ISSUE}

%% ---------------------------------------------------------------------
%% Abstract
%% ---------------------------------------------------------------------
\begin{abstract}%
    \input{content/00_abstract}

\end{abstract}

%% ---------------------------------------------------------------------
%% Content
%% ---------------------------------------------------------------------

\maketitle

\input{content/10_introduction}
\input{content/30_ambientanalytics}

\input{content/50_examples}
\input{content/60_challenges}

\input{content/99_conclusion}

\section{ACKNOWLEDGMENTS}

We thank the Alexander von Humboldt Foundation funded by the German Federal Ministry of Education and Research.
This research was partially supported by Villum Investigator grant VL-54492 by Villum Fonden and the Deutsche Forschungsgemeinschaft (DFG, German Research Foundation) under Germany's Excellence Strategy -- EXC 2120/1 -- 390831618

%% ---------------------------------------------------------------------
%% References
%% ---------------------------------------------------------------------

\setcounter{enumiv}{0}
\bibliographystyle{abbrv}
\bibliography{references}

%% ---------------------------------------------------------------------
%% Biographies
%% ---------------------------------------------------------------------

\begin{IEEEbiography}{Sebastian Hubenschmid}{\,} is a postdoctoral research fellow at the Department of Computer Science at Aarhus University, Denmark.
% His research combines human-computer interaction, mixed reality, and data visualization.
He received his Ph.D. in computer science from the University of Konstanz, Germany.
He is a member of the IEEE.
Contact him at ssh@cs.au.dk.
\end{IEEEbiography}

\begin{IEEEbiography}{Arvind Srinivasan}{\,} is a Ph.D.\ candidate %and Research Fellow
at the Department of Computer Science at Aarhus University, Denmark.
He earned his Masters' in Human-Computer Interaction from the University of Maryland, College Park.
%His research combines context-aware systems, immersive analytics, and data visualization. He is a member of the IEEE. Contact him at arvind@cs.au.dk.
\end{IEEEbiography}

\begin{IEEEbiography}{Niklas Elmqvist}{\,} is a Villum Investigator and professor in the Department of Computer Science at Aarhus University in Aarhus, Denmark.
%His research interests include data visualization, human-computer interaction, and human-centered AI.
He received the Ph.D. degree in computer science from Chalmers University of Technology.
He is a Fellow of the IEEE and the ACM.
Contact him at elm@cs.au.dk.
\end{IEEEbiography}

\begin{IEEEbiography}{Dieter Schmalstieg}{\,}  is Alexander von Humboldt Professor of Visual Computing at the University of Stuttgart, Germany.
His research interests span augmented reality, virtual reality, and visualization.
% Ph.D.?
%He is a Fellow of the IEEE and the IEEE Computer Society.
Contact him at schmaldr@visus.uni-stuttgart.de.
\end{IEEEbiography}

\begin{IEEEbiography}{Michael Sedlmair}{\,} is a full professor at the University of Stuttgart, Germany, where he leads the Human-Computer Interaction research group.
%where he is leading the Visualization and Virtual/Augmented Reality research group.
He received the Ph.D. degree in computer science from the University of Munich, Germany.
%%His research interests include visual and interactive machine learning, perceptual modeling for visualization, immersive analytics and situated visualization, novel interaction technologies, as well as the methodological and theoretical frameworks underlying them.
He is a member of the IEEE and ACM.
Contact him at michael.sedlmair@visus.uni-stuttgart.de.
\end{IEEEbiography}

\end{document}

%% file: commands.tex
\newcommand{\lastaccess}{2025-12-16}

%% file: content/00_abstract.tex
% !TEX root = ../main.tex
\hl{Augmented} reality has great potential for embedding data visualizations in the world around the user.
While this can enhance users' understanding of their surroundings, it also bears the risk of overwhelming their senses with a barrage of information.
In contrast, calm technologies aim to place information in the user's attentional periphery, minimizing cognitive load instead of demanding focused engagement.
In this column, we explore how visualizations can be harmoniously integrated into our everyday life through \hl{augmented} reality, progressing from visual analytics to \emph{ambient analytics}.

%% file: content/10_introduction.tex
% !TEX root = ../main.tex

\chapteri{H}ow will our world change if we can effortlessly infuse reality with digital information?
\hl{Using ``augmented reality''~(AR)} for continuous access to information---represented as visualizations, not text---is especially compelling as it promises to support ubiquitous analytics~\cite{elmqvist2013ubiquitous}.
Contemporary mobile devices, such as smartphones or tablets, cannot truly be used anytime and anywhere.
They typically occupy both hands and monopolize the user's attention, making them unsuitable for crowded or rapidly changing environments, and, at least, awkward in social situations.
With the advent of modern \hl{AR} headsets and ``smartglasses'' of the near future, continuous access to visualizations is becoming more practical. 
However, na\"ively migrating apps from the smartphone ecosystem to floating windows displayed via smartglasses, while replacing touchscreen operation with mid-air gestures, is unlikely to appeal to large audiences. 

\begin{figure}
    \centerline{\includegraphics[width=\columnwidth]{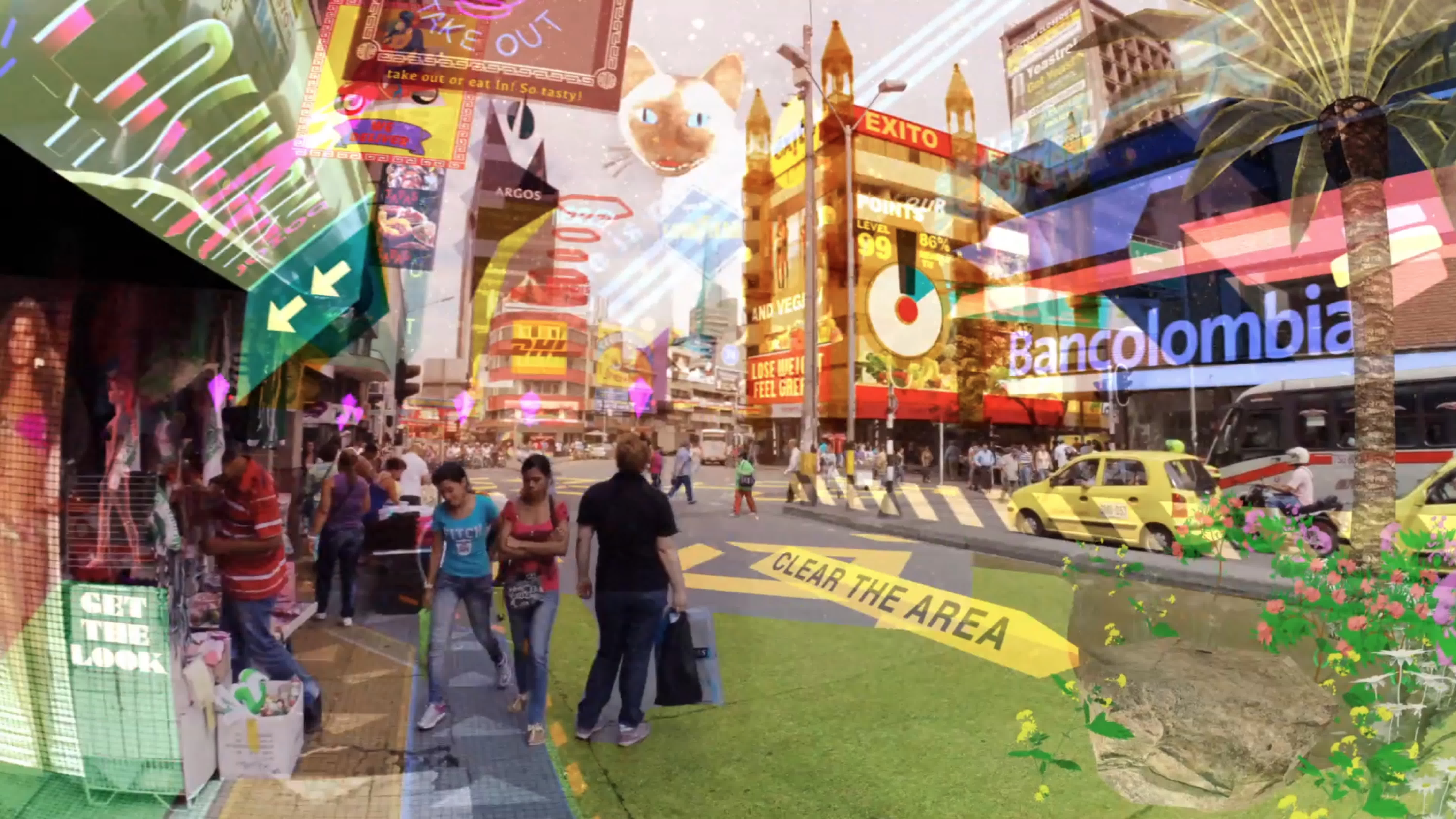}}
    \caption{%
    A future for ubiquitous visualizations supported through smart glasses may quickly lead to a \hl{cacophony} of attention-grabbing notifications and visualizations.
    Image source: Keiichi Matsuda, Hyper-Reality,\protect\footnotemark{} used with permission.
    }
    \label{fig:teaser}
\end{figure}

\hl{Yet, in current practice, this is the trajectory often followed when adopting AR technology.}
\hl{Immersive and situated analytics}~\cite{kraus2021value} often suggest such a future of salient visualizations \hl{at the forefront of our field of view}.
Following best practices in information clarity, \hl{visualizations are rendered so they stand out from their environment, or are} prominently overlaid on top of everyday objects.
As a result, these visualizations are designed to fully monopolize the user's attention---a resource that is becoming increasingly scarce.
Will future AR environments be a cluttered \hl{dystopia} full of attention-grabbing notifications (\autoref{fig:teaser}), or could we instead use these newfound capabilities to integrate visualizations harmoniously into our environment?

This dichotomy is the premise of \textit{calm technology}~\cite{weiser1997coming}, which has been applied to the field of visualizations as \textit{ambient information systems}~\cite{pousman2006taxonomy}.
The goal is to create visualizations that reside in the user's attentional periphery, seamlessly integrated with their environment, thus minimizing cognitive load.
Yet, prior work in this space has largely been constrained by technical limitations, restricting these visualizations to the confines of 2D displays.
With the increasing viability of AR, as well as the growing capabilities of artificial intelligence~(AI), we see untapped potential that extends beyond the confines of conventional displays and leverages ongoing developments in immersive analytics~\cite{kraus2021value}.
These capabilities open up a new design space that allows us to integrate data visualizations as natural phenomena~\cite{kouts2023lsdvis} or deeply embed visualizations into everyday objects.

In this column, we extrapolate current hardware trends towards a hypothetical ultimate display~\cite{sutherland1965ultimate} that supports the harmonious integration of data visualizations into the user's surroundings.
We introduce the term \textit{ambient analytics} to describe the process of subliminal sensemaking that emerges from perceiving and interacting with such ambient visualizations across different modalities.
We highlight emerging use cases of ambient visualizations and outline the opportunities and challenges, paving the way for a research agenda in this nascent field.

\footnotetext{\href{http://hyper-reality.co}{http://hyper-reality.co}, last accessed \lastaccess{}.}

%% file: content/30_ambientanalytics.tex
% !TEX root = ../main.tex

\begin{mdframed}[backgroundcolor=blue!10,
    frametitle={\section*{\textcolor{white}{Background}}},
    frametitlerule=true, frametitlebackgroundcolor=bgcolor]

The concept of visualizing data anytime and anywhere using the possibilities of modern devices has been previously explored in prior work~\citebg{Elmqvist2023}.

\paragraph{Ubiquitous Analytics}

The widespread availability of screens, mobile devices, and, increasingly, \hl{AR} headsets allows us to integrate interactive digital media into every fabric of our lives.
This \textit{ubiquitous} availability of technology enables a ``deep and dynamic analysis of [...] data anytime and anywhere''~\cite{elmqvist2013ubiquitous}.
\textit{Ubiquitous analytics} thus focuses on the potential of analytics beyond the desktop, unchaining visualizations from the confines of our desks.
However, while this direction focuses on the reachability and availability, it makes no distinction in \textit{how} to integrate these visualizations into our world.

\paragraph{Pervasive Augmented Reality}

Closely related to \textit{ubiquitous analytics} are \textit{pervasive computing} and \textit{pervasive augmented reality}~\citebg{grubert2017pervasive}.
Whereas \textit{ubiquitous analytics} focuses on the possibilities of making visualizations available anytime and anywhere, pervasive technologies are often concerned with adapting technology to the user's current context, providing a ``continuous and multi-purpose user experience''~\citebg{grubert2017pervasive}.
Such pervasive technologies thus can inform longitudinal adaptations of visualizations in our world.

\paragraph{Situated and Embedded Visualizations}

The combination of \textit{pervasive displays} and \textit{ubiquitous analytics} allows us to visualize and analyze data \textit{anywhere}.
Building on this idea, situated analytics describes the process of sensemaking where both the origin of data (i.e., the physical referent) and its visualization are closely temporally and spatially connected~\citebg{thomas2018situated}.
This form of sensemaking is inherently local, with supporting visualizations ranging from merely situated to truly embedded~\citebg{willett2017embedded}.
Ambient visualizations, in contrast, may be highly situated within the user's environment.
However, we see their potential in the broader context of supporting users through embedded visualizations in tasks that may or may not be situated.

{\small
\makeatletter
\renewcommand{\@biblabel}[1]{[B#1]}
\makeatother

\bibliographystylebg{abbrv}
\bibliographybg{references}
}

\end{mdframed}

\section{TOWARDS AMBIENT ANALYTICS}

How can we integrate visualizations anytime and anywhere without overwhelming the user with a barrage of salient visualizations?
We think that this sensemaking process could be supported by environmental cues that reside in the periphery of attention.
Such \textit{ambient visualizations}~\cite{pousman2006taxonomy} leverage \textit{calm technologies}~\cite{weiser1997coming} to avoid a conflict with the user's environment.
To account for recent \hl{emergent technologies}, we broadly define ambient visualizations as follows:

\begin{quote}
    Ambient visualizations are low-saliency and highly embedded in the user's environment, thus residing in the periphery of the user's attention.
\end{quote}

Immersive analytics opens up entirely new ways of integrating ambient visualizations into the environment~\cite{kouts2023lsdvis}.
Research in this area has broadened the design space beyond purely visual cues, enabling information to be conveyed through auditory or olfactory channels.
Exposure to such multisensory cues could enable a gradual sensemaking process in which users form understanding over time.
Hence, to move towards such a holistic \hl{analysis} process from ambient cues, we \hl{see the potential for } \textit{ambient analytics} in the following sense:

\begin{quote}
    Ambient analytics \hl{supports the formation of} intuition and subliminal sensemaking through the multisensory perception of intentional environmental cues, such as ambient visualizations.
\end{quote}

\hl{
At this point, we must acknowledge the tension between \textit{ambient}---typically meaning implicit---and \textit{analytics}, commonly understood as explicit.
Prior work in problem solving suggests that insight arises not only through explicit thinking, but that it also requires unconscious analytical thought processes}~\cite{macchi2019insight}.
\hl{
Consequently, our notion is not intended to prescribe, but rather to foment new research on how emergent technologies, such as immersive and situated analytics, might support conventional analytics through implicit rather than explicit means.
}

\subsection{Design Dimensions}

\begin{figure*}
    \centering
    \includegraphics[width=0.72\textwidth]{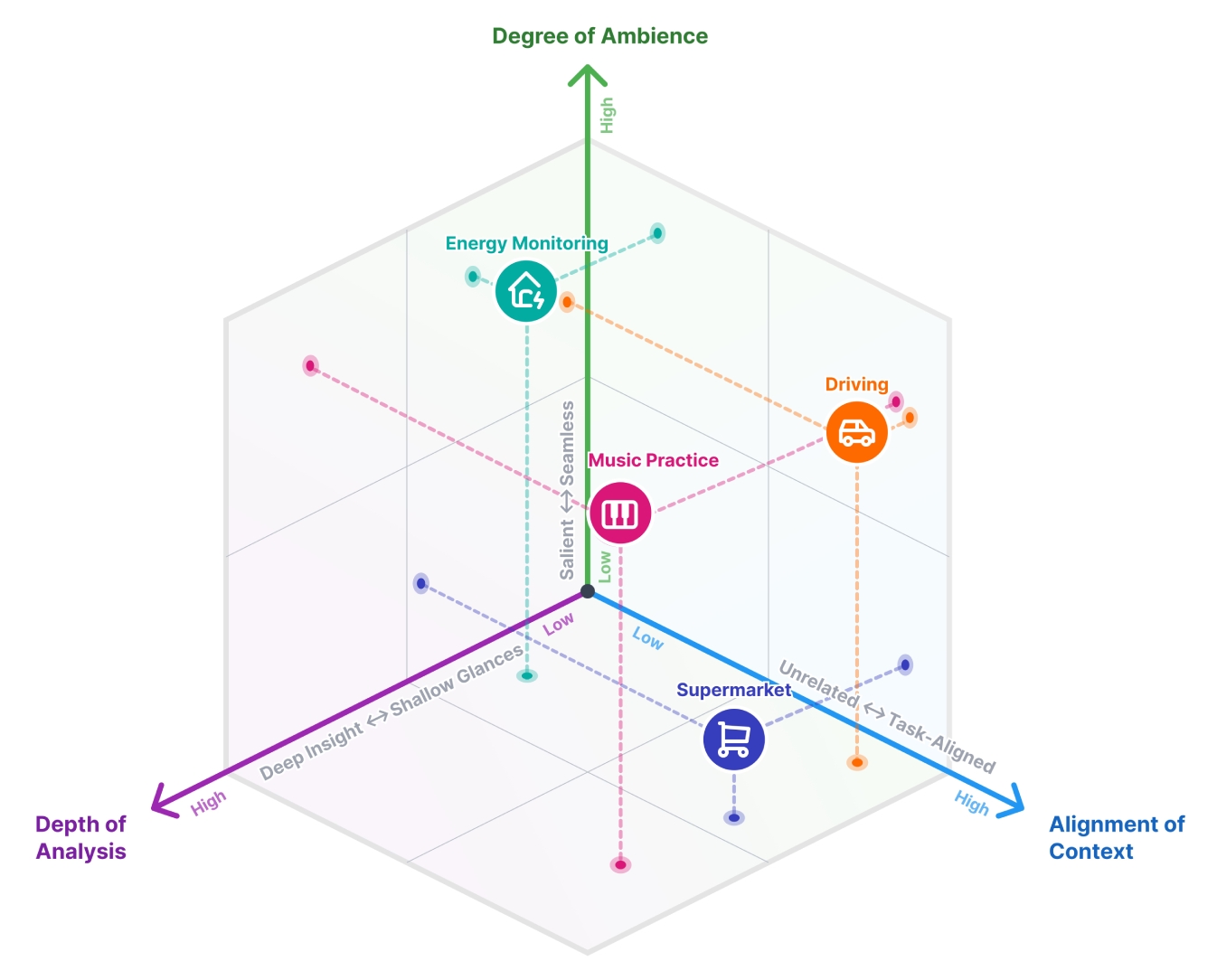}
    \caption{%
    Three design dimensions for ambient analytics systems, illustrated with four scenarios.
    \textit{Degree of Ambience} (vertical) captures how seamlessly the visualization integrates into its surroundings.
    \textit{Alignment of Context} captures relevance to the user's primary task.
    \textit{Depth of Analysis} captures the extent to which the sensemaking process is supported.
    Projection lines indicate each scenario's position along each axis.
    }
    \label{fig:design-dimensions}
\end{figure*}

We propose three \hl{initial} design dimensions to \hl{characterize} ambient analytics systems (\autoref{fig:design-dimensions}).
\hl{We see these dimensions as starting points for a more exhaustive exploration of this space, which may further draw from adjacent taxonomies}~\cite{bartram2015design}.

\paragraph{Degree of Ambience}

The \textit{degree of ambience} refers to the extent to which the sensemaking process disrupts the user's primary activity, taking up mental resources that would otherwise be dedicated to the primary task.
This can range from the visualization itself (i.e., how seamlessly the visualization is integrated into its surroundings) and the amount of mediation of reality required, to the mental workload required to derive insights.
A high degree of ambience may improve aesthetics but also increase workload required for extracting insights, especially if a great \textit{depth of analysis} is needed.
Ambient visualizations should therefore balance informativeness with low visual complexity.

\paragraph{Alignment of Context}

The \textit{alignment of context} refers to the extent to which the ambient sensemaking process is relevant to the user's primary activity.
Since an ambient sensemaking process may occur subliminally, it can often foster intuition about tasks unrelated to the user's current activity. 
However, it may also support the user's primary activity through multi-sensory enrichment and contribute to a deeper understanding.

\paragraph{Depth of Analysis}

As ambient visualizations are intentionally positioned at the periphery of attention, \hl{they were often limited to convey} non-critical insights that ``support mundane activities''~\cite{skog2003aesthetics}.
\hl{However,} the \textit{depth of analysis} can vary between different systems and, thus, refers to how much of the sensemaking process is supported by digital cues.
A greater depth of analysis may, over time, foster deeper insights \hl{and potentially lead to immediate or near-term decisions that are important,} or invite interactive exploration for further investigation.
In contrast, a shallow depth of analysis may yield only superficial, high-level insights.

%% file: content/50_examples.tex
% !TEX root = ../main.tex

\section{ILLUSTRATIVE SCENARIOS}
In this section, we explore hypothetical examples of ambient analytics systems based on prior work in the area of ambient visualizations.

\subsection{Passive Data Awareness}

Although we are surrounded by a plethora of data, we are often only interested in data deviating from the norm.
Take, for example, a residential use case: 
Our increasingly smart homes enable sophisticated insights into energy usage or heating.
Assuming that everything is operating normally, users may not want to fill their homes with conventional data visualizations about their energy usage.
Defining alarms when energy usage becomes too high---and ensuring that the alarm system works properly---may be beyond the technical capabilities of the average user, especially since such feedback systems can vary widely depending on external factors.

\begin{figure}
    \includegraphics[width=\columnwidth]{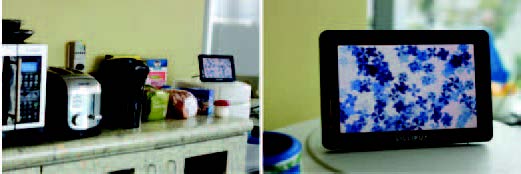}
    \caption{Ambient displays can be used to depict energy usage through artistic renditions. Image source: Rodgers et al.~\cite{rodgers2011exploring}, used with permission.}
    \label{fig:energy}
\end{figure}

In this context, prior work~\cite{rodgers2011exploring} has employed abstract visualizations that depict energy usage through artistic renditions on ambient displays (\autoref{fig:energy}).
With the ultimate display, we would no longer be confined to ambient screens and could instead digitally mediate the perception of our homes to visualize this data.
For example, generative AI can be used to alter existing wall art~\cite{skog2003aesthetics}, integrating data visualizations that mimic natural phenomena~\cite{kouts2023lsdvis} or modifying the image to reflect some data (e.g., displaying a weather forecast) while preserving most of the original artwork.
Relevant architectural necessities, such as light that leaks through door gaps, could be used to encode how much heat has escaped from this room over the past day.
In addition, sensory cues, such as sounds and scents, can be gently embedded into users' homes, providing a subtle, yet ever-present awareness of the environmental data.

Such \textit{passive data awareness} prioritizes a harmonious integration into the user's home (high degree of ambience) over in-depth insights (low depth of analysis).
The integration in this scenario is intentionally easy to overlook, blending aesthetically into the background of a user's home.
Previously hidden data is brought to the user's awareness without intruding on their everyday tasks (low contextual alignment).
Instead, it could serve as a starting point for deeper analysis when complemented by on-demand interactive views using more conventional visualizations.

\subsection{Calm Environmental Embedding}

The promise of visualizing data anywhere and anytime often clashes with the messy reality of the outside world.
In supermarkets, for example, shoppers want to make informed decisions, but are hindered by a multitude of external factors, such as adversarial product placement, marketing tactics, and other customers.
How can we support users in such environments without being drowned out by the noise?

Using an ultimate display, we can mediate reality to encode information within existing perceptual noise \hl{based on user preference and intention}.
While altering physical geometry can be dangerous (users must still navigate physical shelves), ambient cues can safely modify color and contrast to convey information about different products.
Additionally, the ultimate display can embed visualizations into the physical environment, such as shelves and product labels.
Here, cues may need to be more pronounced to avoid being overshadowed by external stimuli.

A higher salience counteracts a noisy environment (low degree of ambience) and supports the users in their task (high contextual alignment).
Given the challenges of interacting with digital information in public spaces, such an ambient system may already provide a greater depth of analysis, reducing the need for fully interactive systems.

\subsection{Attention-Preserving Cues}

Ambient visualizations can also provide supplemental information during primary activities, where the analytical needs are reduced in favor of lower cognitive load.
One example is driving a car: any visualization that competes with the primary task could distract the driver and thus needs careful consideration.
While primary tasks are supported by salient information displays such as speedometers, further assisting drivers by conveying information about energy consumption or fuel efficiency can be desirable.

\begin{figure}
    \includegraphics[width=\columnwidth]
    {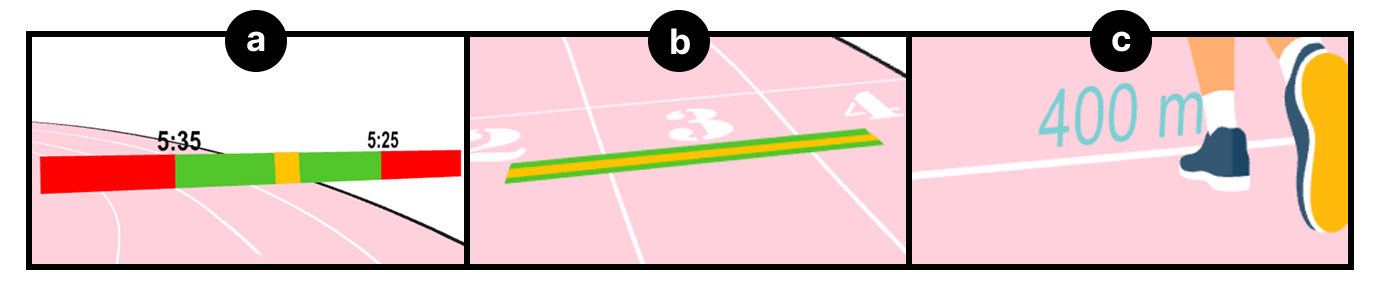}
    \caption{
        Situated visualization on a track for conveying pacing information to runners.
        The study found all three \hl{situated} visualizations (a.~\textit{Speedometer}, b.~\textit{LaserBeam}, c.~\textit{ShrinkingLines}) resulted in better or similar pace regularity compared to the traditional smartwatch methods.
        Image source: Li et al.~\cite{li2025embedded}, used with permission.
    }
    \label{fig:running}
\end{figure}

To keep the user focused on the primary task, ambient analytics may rely on environmental cues.
We consider the subtle enrichment of elements already within the user's focus, such as \hl{road or lane markings}, to convey additional information without distraction (\autoref{fig:running}).

Unlike other ambient visualizations that prioritize aesthetic integration into the user's surroundings, this scenario may benefit from a clearer representation to reduce the mental effort required for interpreting the visualization.
However, the degree of ambience should be high to embed this information into the user's environment.
A high contextual alignment is necessary, as all other information should be subdued to avoid distractions.
However, a low depth of analysis is appropriate to avoid masking critical information.

\subsection{Fostering Intuition}

\begin{figure}
    \includegraphics[width=\columnwidth]{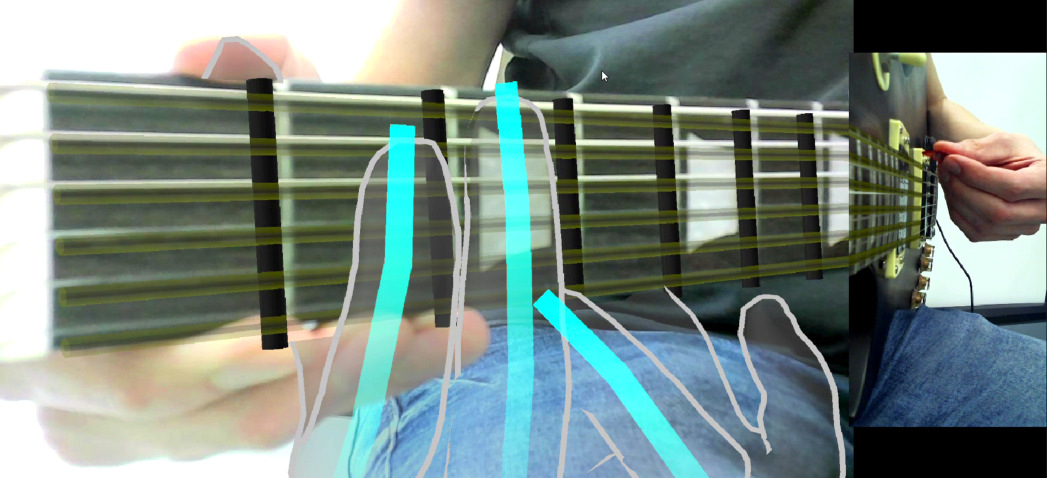}
    \caption{
        A study on AR guitar tutorials using fret highlighting found that simpler visualizations were preferred over detailed hand overlays, suggesting that reduced saliency can improve learning for fine-grained tasks.
        Image source: Skreinig et al.~\cite{skreinig2023guitarhero}, used with permission.
    }
    \label{fig:guitar}
\end{figure}

Ambient visualizations can support the gradual development of expertise over time.
Consider a student practicing a musical instrument: Traditional instruction either lacks awareness of attention breakdowns or introduces overly explicit visualizations that disrupt musical flow.
Recent work on AR guitar tutorials~\cite{skreinig2023guitarhero} illustrates this tension (\autoref{fig:guitar}), finding that users prefer simpler fret highlights over detailed virtual hand overlays, as the latter introduce visual clutter.

Attention-aware visualizations~\cite{srinivasan2024attention} combined with diminished reality could extend such approaches toward a more ambient alternative.
By reducing the salience of environmental distractions and dimming non-relevant elements, the system blends into the practice environment while scaffolding the development of expert gaze patterns through repeated exposure rather than explicit instruction.

This support must adapt as expertise develops, since guidance that is beneficial for novices can become annoying for advanced learners.
An adaptive ambient system should use stronger diminished reality for novices and progressively relax as expertise grows.
This scenario exemplifies both a high degree of ambience and a great depth of analysis, while maintaining high contextual alignment by directly supporting musical practice.

%% file: content/60_challenges.tex
% !TEX root = ../main.tex

\section{OPPORTUNITIES AND CHALLENGES}

We describe the challenges and opportunities associated with the design, implementation, and evaluation of ambient analytics systems.

\subsection{Building Intuition}

One of the primary benefits of ambient visualizations lies in their non-intrusive yet persistent presence.
Unlike conventional visualizations that demand focused attention, ambient visualizations provide continuous exposure to data over extended periods.
This sustained, low-intensity engagement can foster intuition grounded in factual data rather than relying solely on trial and error---a form of implicit learning where knowledge is acquired through exposure without conscious awareness~\cite{Reber1989Implicit}.
For instance, a trader who peripherally observes ambient representations of market volatility over weeks may develop an intuitive sense for market rhythms---recognizing patterns that feel ``off'' before being able to articulate why.
However, empirical validation of such intuition-building effects remains an open challenge, requiring longitudinal studies that track how ambient exposure shapes mental models over time.

\subsection{Guidelines}

Conventional visualization guidelines emphasize the clear conveyance of information, clashing with the goals of ambient visualizations that often favor aesthetic integration.
\hl{Visualizations in dashboards, for example, aim for precision when visualizations are viewed in isolation, but collectively exhibit ambient qualities, such as glanceability, that are traded off against precision}~\cite{sarikaya2019what}.
\hl{This raises the question: How much of what this work described as ``ambient'' is already hidden in existing visualizations?}
Additionally, mixed reality enables the unprecedented integration of visualizations into the environment, which goes beyond existing 2D ambient visualizations.
We must thus carefully re-examine existing guidelines regarding these ambient visualizations and establish new guidelines to support their design. 

\begin{figure*}
    \includegraphics[width=\textwidth]{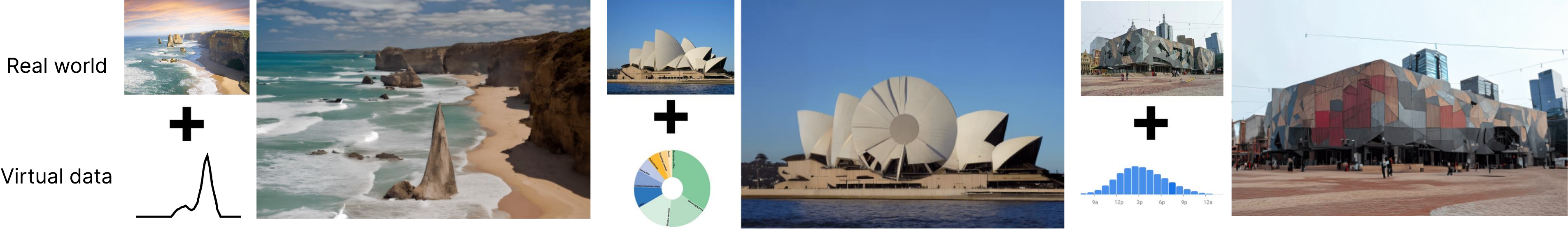}
    \caption{
    Generative artificial intelligence can be used to embed visualizations as natural phenomena into the environment.
    From left to right: a rock formation encodes a line chart in a coastal landscape, the Sydney Opera House shells transform into a pie chart, and Federation Square's facade displays a bar chart.
    Image source: Kouts et al.~\cite{kouts2023lsdvis}, used with permission.
    }
    \label{fig:lsdvis}
\end{figure*}

\subsection{Interactivity}

To drive the process of subconscious sensemaking in ambient analytics, we must also consider some form of interaction with these visualizations.
The design space of immersive analytics provides a few opportunities:
For example, systems can sense the general engagement with the ambient visualization (e.g., using gaze data) and transition from an ambient visualization to a fully interactive visualization.
Alternatively, smartglasses can infer the user's intentions in the physical world and adapt the ambient visualization implicitly, keeping the visualization within the attention periphery.
However, mediating the user's environment (\autoref{fig:lsdvis}) could introduce additional challenges, as the illusion may break when users want to physically interact with digitally mediated objects---or ``\textit{data morganas}''---unless you have appropriate feedback mechanisms.

\subsection{Literacy}

How can we ensure that users understand the visualization, given the wide range of literacy levels in the visualization domain?
Ambient visualization should be well-integrated into the user's environment, favoring aesthetics over clarity~\cite{skog2003aesthetics}.
However, this choice can quickly make visualizations only decipherable by experts.
Mixed reality opens up new possibilities for appropriating the environment for ambient visualizations~\cite{kouts2023lsdvis}, but it also bears the risk of accidentally interpreting naturally occurring patterns as intentional visualizations (\autoref{fig:lsdvis}).
Thus, users must be able to clearly differentiate between digitally mediated information and reality, while the system should adapt to the user's visualization literacy.
\hl{Here, we also see the potential in leveraging AI to tailor visualizations to the intents of the user, support their decision-making, and learn their behavior over time.}

\subsection{Implementation}

In this column, we extrapolated from current trends and assumed that users have the ultimate display that enables them to fully mediate their own reality.
However, we are still far from such a hypothetical device.
How can we realize ambient analytics using current technology?

On the one hand, smart devices with output capabilities (e.g., lights) are still becoming increasingly ubiquitous, and screens can be integrated seamlessly, for example, using e-ink technology.
Although most screens are designed for ``single focus'' use, they could also be appropriated as ambient displays when not in use.
AI could help us find the current focus of use and subtly display relevant information using surrounding devices.
While this bears the risk of overloading the user's environment with unwanted information, it also has the potential for more adaptive interfaces.

On the other hand, smartglasses for everyday use are entering the consumer market, enabling data visualization anywhere, anytime.
With state-of-the-art computer vision methods, we can already convincingly embed data visualizations into static environments (\autoref{fig:lsdvis}).
While we may be far from any consumer-ready ambient visualizations, current technology can already serve as a testbed to investigate how to design and evaluate ambient visualizations.

\subsection{Evaluation}

A good ambient visualization fades into the background, supporting the user through subconscious sensemaking.
This makes evaluating such workflows especially challenging, as we need to measure not only the amount of information conveyed but also how distracting or intrusive the visualization is.
Especially for aspects such as building intuition, longitudinal methods are required, which have historically been underrepresented in the fields of human-computer interaction and visualization.

%% file: content/99_conclusion.tex
% !TEX root = ../main.tex

\section{CONCLUSION}

Recent advances in mixed reality and artificial intelligence bring us closer to Sutherland's vision of the ultimate display---allowing us to seamlessly mediate the perception of our physical world with based on digital information.
While this trend opens new possibilities for analyzing data anytime and anywhere, it also risks overwhelming users.

As a counterbalance, we revisit the concept of calm technology and propose ambient visualizations in the context of the theoretical ultimate display.
We derive the notion of \textit{ambient analytics} to describe the process of subconscious sensemaking through ambient visualizations and digitally mediated environmental cues.
In contrast to conventional data analysis, which relies on salient visualizations and sustained focus, our notion of ambient analytics emphasizes intuition-building through harmoniously embedded representations.
We thus see the potential to explore this nascent field further, ensuring that future technologies help build intuition instead of contributing to information overload.

%% file: ambient.bbl
\begin{thebibliography}{1}

\bibitem{Elmqvist2023}
N.~Elmqvist.
\newblock Data analytics anywhere and everywhere.
\newblock {\em Communications of the ACM}, 66(12):52--63, 2023.

\bibitem{grubert2017pervasive}
J.~Grubert, T.~Langlotz, S.~Zollmann, and H.~Regenbrecht.
\newblock Towards {{Pervasive Augmented Reality}}: {{Context-Awareness}} in {{Augmented Reality}}.
\newblock {\em IEEE Transactions on Visualization and Computer Graphics}, 23(6):1706--1724, June 2017.

\bibitem{thomas2018situated}
B.~H. Thomas, G.~F. Welch, P.~Dragicevic, N.~Elmqvist, P.~Irani, Y.~Jansen, D.~Schmalstieg, A.~Tabard, N.~A.~M. ElSayed, R.~T. Smith, and W.~Willett.
\newblock Situated {{Analytics}}.
\newblock In {\em Immersive {{Analytics}}}, volume 11190, pages 185--220. Springer Publishing, Cham, 2018.

\bibitem{willett2017embedded}
W.~Willett, Y.~Jansen, and P.~Dragicevic.
\newblock Embedded {{Data Representations}}.
\newblock {\em IEEE Transactions on Visualization and Computer Graphics}, 23(1):461--470, Jan. 2017.

\end{thebibliography}
